\documentclass[aip,pop,reprint]{revtex4-1}
\usepackage{graphicx}
\usepackage{dcolumn}
\usepackage{bm}

\begin{document}

\title{Particle conservation in numerical models of the tokamak plasma edge}
       
\author{Vladislav Kotov}

\email{v.kotov@fz-juelich.de}

\affiliation{Forschungszentrum J\"ulich GmbH, Institut f\"ur Energie- und Klimaforschung - Plasmaphysik,  
Partner of the Trilateral Euregio Cluster (TEC), 52425 J\"uich, Germany}

\begin{abstract}
The test particle Monte-Carlo models for neutral particles are often used 
in the tokamak edge modelling codes. The drawback of this approach is that  
the self-consistent solution suffers from random error introduced by the statistical method.
A particular case where the onset of nonphysical solutions can be clearly identified 
is violation of the global particle balance due to non-converged residuals. 
There are techniques which can reduce the residuals - such as internal iterations in the code B2-EIRENE - 
but they may pose severe restrictions on the time-step and slow down the computations.
Numerical diagnostics described in the paper can be used to unambiguously identify when the too large 
error in the global particle balance is due to finite-volume residuals, and 
their reduction is absolutely necessary. Algorithms which reduce the error while allowing large time-step are also discussed. 
\end{abstract}

\maketitle

\section{Introduction}

A combination of a 2D finite-volume plasma transport code with a 
kinetic Monte-Carlo model for neutral particles is typically applied for numerical modelling of the 
tokamak edge and divertor plasmas. A well known example of such modelling tool is 
the code package B2-EIRENE~\cite{ReiterJNM1992,Reiter_EIRENE2005} (SOLPS)
widely used in the field. The Monte-Carlo method allows physically 
accurate description of atomic and molecular kinetics in complex geometries, but 
has a disadvantage of random error - statistical noise in the calculated quantity. 
There were always concerns that 
this statistical noise can have detrimental impact on the coupled solution~\cite{Maddison1994}. 

In the present paper one specific noise related issue 
which can lead to pathological solutions is addressed - violation of the global particle balance. 
It is shown that the error in the steady-state particle balance can be 
presented as a sum of three terms. Those are the operator splitting error, 
residual of the fluid solver, and the time-derivative. 
Whereas the first term can be effectively reduced by the source re-scaling, the reduction of residuals 
may require iterative solution of the discretized fluid equations after each call of the Monte-Carlo model. 
This can, in turn, pose severe restrictions on the time-step and lead to a very long overall run-time. E.g. in the ITER modelling studies~\cite{KukushaFED2011}
one model run could take several months of wall-clock time. 
Special diagnostics for monitoring of the particle balance allow to clearly identify the 
cases when reduction of residuals is absolutely necessary, and the corresponding measures 
must be taken. 

This paper presents in condensed form  the  most important findings from the dedicated 
studiy of the SOLPS code~\cite{KotovJul2014}. 
Prototypes of the numerical diagnostics were implemented and tested 
in the code SOLPS4.3 which is the legacy version of B2-EIRENE used in the past for 
the ITER design modelling~\cite{KukushaFED2011}.
The approach itself is thought to be applicable to any 
finite-volume edge code.
The numerical convergence is analyzed here only in terms of the global balances and criteria of the (quasi-)steady-state. 
It is not attempted to use the stricter methods of analysis 
proposed recently for the combination of fluid and Monte-Carlo models in~\cite{GhoosJCP2016}. 
Only steady-state solutions are considered. 

The rest of the paper is organized as follows. In the next section 
a finite-volume fluid code with source terms calculated 
by Monte-Carlo is described in general terms.
In Section~\ref{diagnosticsS} the diagnostics for monitoring of the 
particle balance are introduced.
An example of calculations with different error (residual) reduction 
techniques is discussed in Section~\ref{exampleS}. 
Further methods which can be used to reduce the residuals and the associated error 
in the particle balance 
are outlined in Section~\ref{reductionS}. Last section summaries the conclusions.

\section{Coupling of a finite-volume and a Monte-Carlo models}

Here only minimal information about numerical procedure of the code B2-EIRENE is given which
is required for the subsequent discussion. 
The plasma transport code B2~\cite{BraamsPhDThesis,BraamsNET} solves a set of 2D (axi-symmetric) 
equations for particle conservation, parallel momentum balance, electron and ion energy. 
The full set of equations can be found in Ref.~\onlinecite{BraamsNET}, Chapter 2. The computational domain comprises the 
scrape-off-layer (SOL) region outside of the 1st magnetic separatrix, and the edge of the core 
plasma inside the separatrix.

Finite-volume discretization of the differential equations
~\cite{PatankarBook1980,BraamsNET} leads to a set of algebraic equations which can be symbolically written as:
\begin{equation}
{\rm F}\left( \phi \right) = S\left( \phi \right),\quad \phi=\left\{n_\alpha, u_\alpha, T_e,  T_i \right\}
\label{generalEq}
\end{equation}
Here $\phi$ is the solution vector: $n_\alpha$ is the number density and $u_\alpha$ is the parallel velocity
of the ion fluid $\alpha$, $T_e$ and $T_i$ are the electron and ion temperatures. The discrete variables are defined 
in the cell centers or on the cell faces of the grid. ${\rm F}\left( \phi \right)$ is the non-linear vector function, 
$S\left( \phi \right)$ are the source terms calculated by the test particle Monte-Carlo method in each grid cell.

To find the solution of Equations~(\ref{generalEq}) a 
discrete time-derivative $D$ is added to the equations, and iterations over time are performed. 
On each time-iteration $k$ the solution $\phi_k$ of the following set of equation has to 
be found:
\begin{equation}
{\rm F}\left( \phi_k \right) = \tilde S\left( \phi_{k-1} \right) + D \left(  \phi_k, \phi_{k-1} \right)
\label{timestepEq}
\end{equation}
The ``time derivative'' is defined such that $D\left(  \phi_k, \phi_{k-1} = \phi_k \right) = 0$. 
E.g. for the particle continuity $D \left( n_k, n_{k-1} \right) = \left( n_k - n_{k-1}\right)/\Delta t$, where $\Delta t$ is the time-step.  
The notation with tilde $\tilde S\left( \phi_{k-1} \right)$ underlines that this source term is calculated by Monte-Carlo and 
contains random error, as opposite to the ``exact'' value $S\left( \phi \right)$ which would be 
obtained with the infinite number of test particles. 

In the code B2 the set of non-linear algebraic equations~(\ref{timestepEq}) is solved
by simple iterations and block Gauss-Seidel algorithm (splitting by equations).
The so called ``internal iterations'' of B2 are described in detail in Ref.~\onlinecite{BraamsNET}, Chapter 3, 
one may also refer to Ref.~\onlinecite{KotovJul2014}, Chapter 1.2. 
Approximate solution $\phi^m_k$ obtained at the end of internal iteration $m$ can be inserted 
back into Equation~(\ref{timestepEq}) to find the residual:
\begin{equation}
R = \tilde S\left(\phi^m_k | \phi_{k-1} \right) + D \left(  \phi^m_k, \phi_{k-1} \right)  -  {\rm F}\left( \phi^m_k \right)
\label{residualEq}
\end{equation}
That is, the found $\phi^m_k$ fulfills the equation:
\begin{equation}
{\rm F}\left( \phi^m_k \right) = \tilde S\left(\phi^m_k | \phi_{k-1} \right) + D \left(  \phi^m_k, \phi_{k-1} \right) + R 
\label{internal_iterationEq}
\end{equation}
By comparing with Equation~(\ref{generalEq}) one can see that the difference between $S\left( \phi \right)$ 
and the right hand side of Equation~(\ref{internal_iterationEq}) can be seen as generalization 
of the common residual $R$.  

In the simplest procedure the source terms are calculated 
at the beginning of internal iterations and are fixed afterward. That is,
they stay as $\tilde S\left(\phi_{k-1} \right)$. However, certain modifications
of the sources can be made in the iterative solver to adjust them with the 
changed plasma solution $\phi^m_k$. This modification is reflected in the notation as 
$\tilde S\left(\phi^m_k | \phi_{k-1} \right)$. 

\subsubsection{Measures to ensure particle conservation}

Critical importance of very high accuracy in the global particle 
balances for the reactor-scale edge modelling was recognized 
back at the early stages of the ITER analysis~\cite{KukushaFED2011,KukushaPPCF2002}.
To reach this high accuracy the Monte-Carlo neutral transport code 
must ensure perfect particle conservation in its solution. 
The internal balance in the neutral solver is usually
achieved by re-scaling of the volumetric ion sources estimated by the statistical procedure  
to make them entirely consistent with the primary sources of neutral particles.  
To increase accuracy the particles originating from the 
different primary sources $s$ are sampled independently from each other - 
the source is split into independent ``strata''.
The primary sources of neutrals are: i) recombination of ions on the 
solid surfaces - ``recycling''; ii) volumetric recombination in plasma;
iii) gas puff; iv) erosion. The strength of recycling sources is proportional to 
the ion fluxes.  

If the volumetric ion sources $\tilde S\left(\phi_{k-1} \right)$
stay fixed, but the fluxes of neutralized (recycled) ions change in the course of internal iterations, then 
an imbalance in the sinks and sources occurs. To compensate for this inconsistency 
the sources of ions $\alpha$ coming from recycling strata 
$s$: $\tilde S^s_\alpha\left(\phi_{k-1} \right)$, must be re-scaled as follows:
\begin{equation}
\tilde S^s_\alpha \left( \phi^j_k | \phi_{k-1}\right) =  \frac{Q^s_\beta \left(  \phi^j_k \right) }{Q^s_\beta \left( \phi^0_k  \right)} 
\tilde S^s_\alpha\left( \phi_{k-1}\right)
\label{rescalingEq}
\end{equation}
Here $j$ is the index of internal iteration, $\phi^0_k=\phi_{k-1}$, $Q$ is the total flux of neutralized ions to which the 
source $S^s_\alpha$ is proportional. E.g. if $\alpha$ is He$^+$ then $Q^s_\beta$ is the sum of the fluxes of He$^+$ and He$^{++}$.

\section{Monitoring of the particle balance}

\label{diagnosticsS}

Numerical diagnostic for monitoring of the steady-state global particle balance can be derived from 
Equation~(\ref{internal_iterationEq}) by transforming it into the form:
\begin{eqnarray}
&& {\rm F}\left( \phi^m_k \right) = \tilde S\left(\phi^m_k \right) + \nonumber \\ 
  && + \left[ R +  \tilde S\left(\phi^m_k | \phi_{k-1} \right) - \tilde S\left(\phi^m_k \right) + D \left(  \phi^m_k, \phi_{k-1} \right) \right]
\label{generalized_residualEq}
\end{eqnarray}
Error (inconsistency) in the particle balance is defined separately for each ion species $\beta$. ``Ion species'' here is the chemical element 
as opposite to ``ion fluids'' which are charged states of an element. E.g. species Carbon includes 6 ion fluids from C$^+$ to C$^{6+}$. 

Equation~(\ref{generalized_residualEq}) is applied to the discretized continuity equation for each ion fluid $\alpha$ in each cell $i$. 
Then the sum is calculated:
\begin{widetext}
\begin{equation}
 \sum_i \sum_{\alpha'}\left[ \tilde S^{\alpha'}_i\left(\phi^m_k \right) - {\rm F}^{\alpha'}_i\left( \phi^m_k \right) \right] 
 = \sum_i \sum_{\alpha'}
 \left[ - R^{\alpha'}_i + \tilde S^{\alpha'}_i\left(\phi^m_k \right) - \tilde S^{\alpha'}_i\left(\phi^m_k | \phi_{k-1} \right) 
 - D^{\alpha'}_i\left(  \phi^m_k, \phi_{k-1} \right) \right]
\label{particle_balanceEq}
\end{equation}
\end{widetext}
Here $\sum_i$ is the sum over all grid cells, $\sum_{\alpha'}$ is the sum over all ion fluids which belong to ion species $\beta$.
It is readily seen that zero left hand side of Equation~(\ref{particle_balanceEq}) means perfect balance between 
volumetric sources and fluxes, and the right hand side is the error in the global particle balance of species $\beta$.

Alternative way of writing the particle balance uses formulation via fluxes~\cite{KukushaFED2011,KukushaPPCF2002} :
\begin{equation}
\Delta \Gamma^\beta = \frac{\Gamma^\beta_{puff} + \Gamma^\beta_{core} + \Gamma^\beta_{spt} - \Gamma^\beta_{pump} - \Gamma^\beta_{leak}}
{\Gamma^\beta_{puff} + \Gamma^\beta_{core} + \Gamma^\beta_{spt}}
\label{balance_fluxesEq}
\end{equation}
Here $\Gamma^\beta_{puff}$ is the strength of external particle source - gas puff, $\Gamma^\beta_{core}$ is the ion flux through 
the core grid boundary,  $\Gamma^\beta_{spt}$ is the flux sputtered (eroded) from the solid surfaces, $\Gamma^\beta_{pump}$ is the 
flux (of both ions and neutrals) absorbed on solid surfaces - pumped flux, $\Gamma^\beta_{leak}$ is the flux of atoms which leak to the core. 

The final steady-state solution has to self-adjust in such way that the rate with which the particles 
are removed from the system $\Gamma^\beta_{pump} + \Gamma^\beta_{leak}$ becomes equal to 
the particle input:
\begin{equation}
\Gamma^\beta_{in}=\Gamma^\beta_{puff} + \Gamma^\beta_{core} + \Gamma^\beta_{spt}
\end{equation}
That is, $\Gamma^\beta_{in}$ serves as a scale with which the particle balance error has to be compared. 
The numerical solution can be considered as physically meaningful only if this error $<<\Gamma^\beta_{in}$.

Coming back to Equation~(\ref{particle_balanceEq}), its right hand side yields the following expression for the 
relative error:
\begin{equation}
\Delta^\beta = \Delta^\beta_R +  \Delta^\beta_S + \Delta^\beta_T
\label{balance_residualEq}
\end{equation}
$$
\Delta^\beta_R =   \frac{ - \sum_i \sum_{\alpha'} R^{\alpha'}_i}{\Gamma^\beta_{in}}
$$
$$
\Delta^\beta_S  =
\frac{\sum_i \sum_{\alpha'} \left[ \tilde S^{\alpha'}_i \left(\phi^m_k \right) -
\tilde S^{\alpha'}_i \left(\phi^m_k | \phi_{k-1} \right) \right]}{\Gamma^\beta_{in}}
$$
$$
\Delta^\beta_T  = \frac{1}{ \Gamma^\beta_{in}}\sum_i \sum_{\alpha'} \frac{  n^{\alpha',i}_{k-1} - n^{\alpha',i}_k}{\Delta t}
$$
First term $\Delta^\beta_R$ contains residuals calculated with Equation~(\ref{residualEq}) after the 
end of internal iterations. 
This is the error in the solution of the set of nonlinear finite-volume equations on each time-iteration. 
The term $\Delta^\beta_S$ is due to inconsistency of the neutral-related sources calculated on the ``old'' and ``new'' plasma. 
It can be called an operator splitting error. This term can become large if, e.g., 
the re-scaling procedure, Equation~(\ref{residualEq}), is not implemented. 
Last term $\Delta^\beta_T$ is the time derivative 
which is considered as error when a stationary solution is looked for.

If the plasma fluxes in Equation~(\ref{balance_fluxesEq}) 
are taken from the solution $\phi^m_k$, and the neutral fluxes are calculated on the same 
plasma, then it is easy to show that Equations~(\ref{balance_fluxesEq}) and~(\ref{balance_residualEq}) 
must yield exactly same result when one extra condition is fulfilled. 
This condition is the discrete analogue of the divergence theorem:
\begin{equation}
\sum_i \sum_{\alpha'} {\rm F}^{\alpha'}_i\left( \phi^m_k \right) = \Gamma^{\beta+}_{out} - \Gamma^\beta_{core}
\label{total_fluxEq}
\end{equation}
Here $\Gamma^{\beta+}_{out}$ is the total flux of ions of  species $\beta$ 
to the grid boundaries. The total ion source is calculated as the total source of neutral particles minus their 
pumped and leaked fluxes:
\begin{equation}
 \sum_i \sum_{\alpha'} \tilde S^{\alpha'}_i\left(\phi^m_k \right) 
 =\Gamma^{\beta+}_{out} + \Gamma^{\beta}_{puff} + \Gamma^\beta_{spt} - \Gamma^{\beta}_{pump}  - \Gamma^{\beta}_{leak}
\label{total_sourceEq}
\end{equation}
Volume recombination does not appear in Equation~(\ref{total_sourceEq}) because atoms originating from recombination which 
re-ionize back in plasma do not contribute to the net source, and particles  which are removed from the system are already
included in $\Gamma^{\beta}_{pump}$ and $\Gamma^{\beta}_{leak}$. 
Subtracting Equation~(\ref{total_fluxEq}) from Equation~(\ref{total_sourceEq})
yields the nominator of Equation~(\ref{balance_fluxesEq}).

In practice it makes sense to use both diagnostics in parallel. 
Incorrect particle balance in the solution for neutrals or a mistake in the transfer of ion fluxes to the Monte-Carlo code 
manifests itself as non-physical particle sinks or sources. The diagnostic of Equation~(\ref{balance_residualEq}) may not be able to detect 
them because it does not distinguish between ``legitimate'' and ``illegitimate'' sources and sinks of neutrals. This 
distinction is made in Equation~(\ref{balance_fluxesEq}). The two diagnostics are complimentary to 
each other and enable an additional consistency check. 

\section{An example of case study}

\label{exampleS}

An example discussed here is based on a SOLPS4.3 run 
from the data-base of ITER simulations~\cite{KukushaNF2009}
(case \#1568vk4, see Ref.~\onlinecite{KotovJul2014}, Chapter 4.2). 
The model plasma consists of all charged states of D, He and C.
Power entering the computational domain from the core is equal to 
$P_{SOL}$=80~MW, 47~\% of $P_{SOL}$ is radiated, mainly by C ions. 
The D particle content is 
controlled by the gas puff $\Gamma^D_{puff}$=1.17e22~D-at$\cdot$s$^{-1}$ 
and ion flux from the core $\Gamma^D_{core}$=0.91e22~s$^{-1}$. 
Influx of He ions from the core is set to $\Gamma^{He}_{core}$=2.1e20~s$^{-1}$.
All plasma facing components in the model are covered by carbon. 
The pump is modelled by an absorbing surface in divertor beneath the dome. 
The solution represents a relatively hot attached plasma in front of divertor targets, 
with insignificant parallel momentum losses and volume recombination.

\begin{figure*}
\includegraphics[width=5cm]{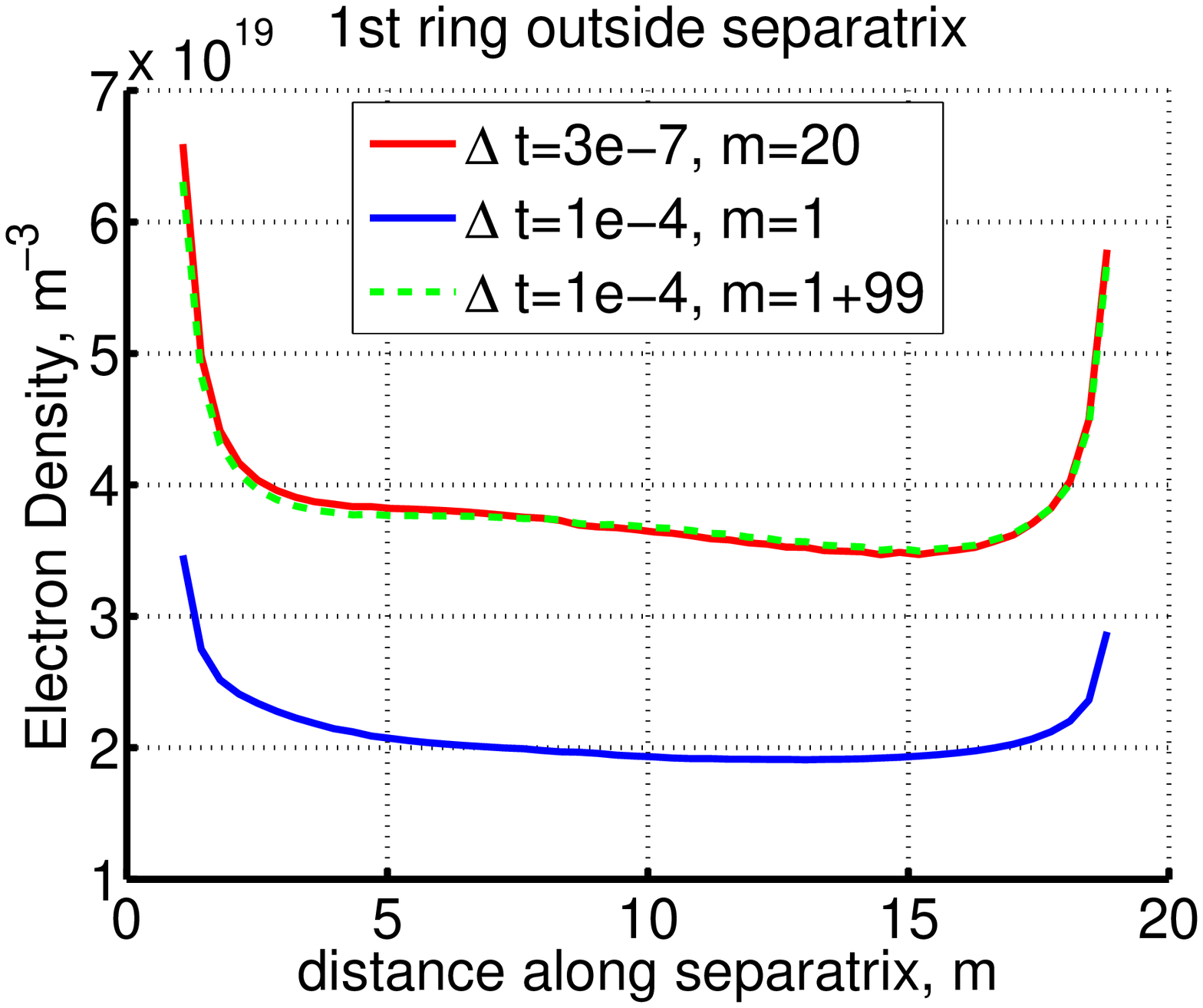}
\includegraphics[width=5cm]{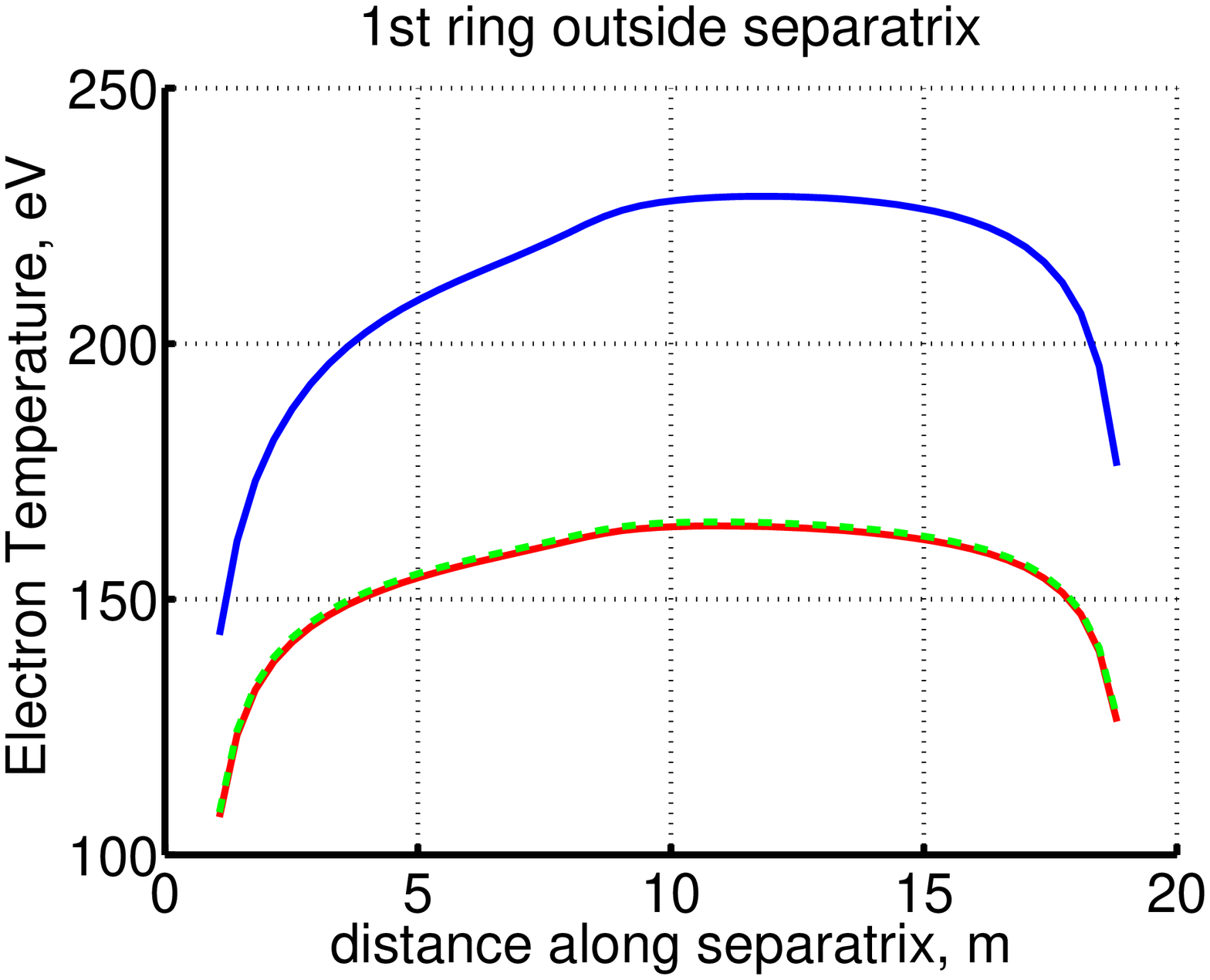}
\includegraphics[width=5cm]{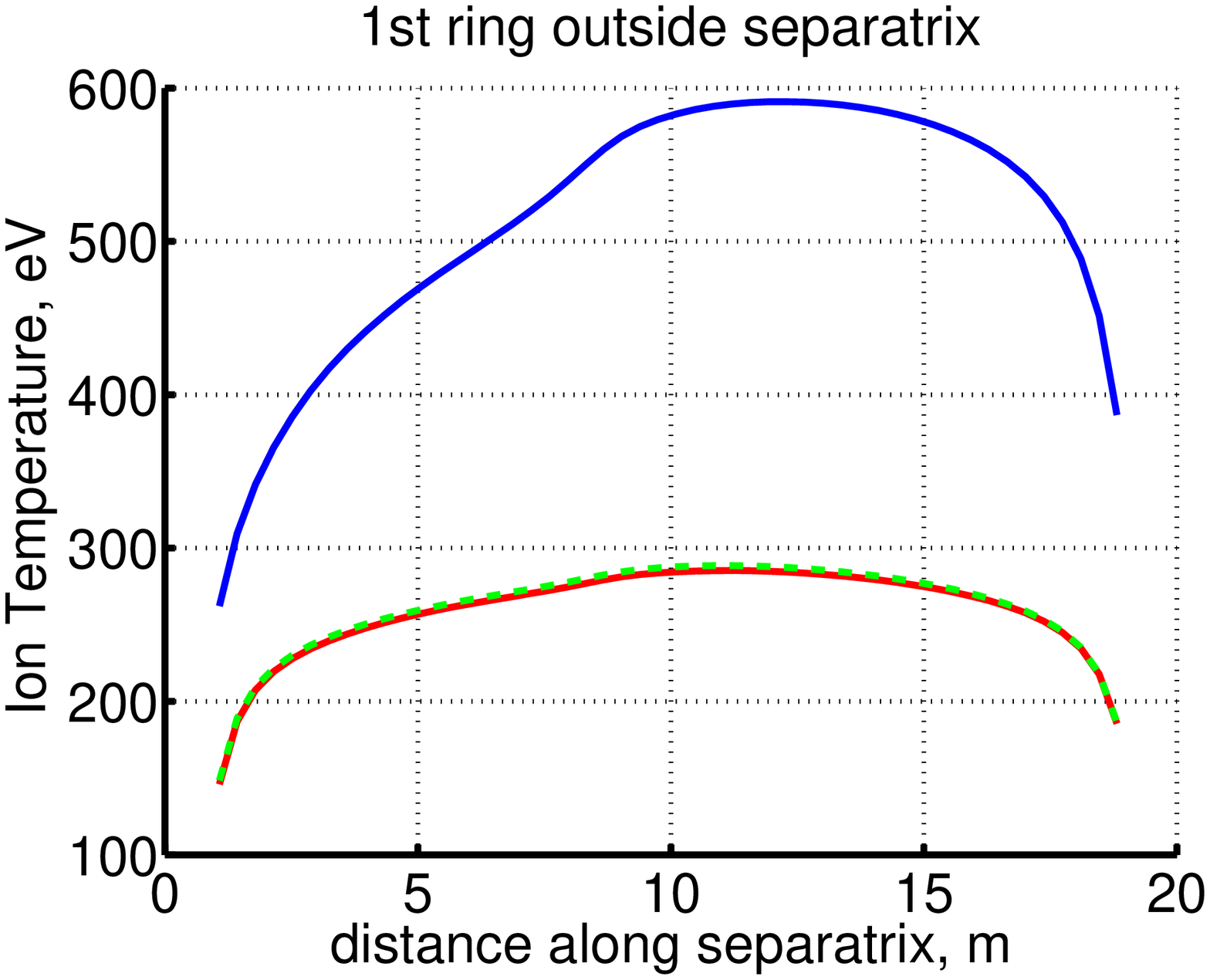} \\
\includegraphics[width=5cm]{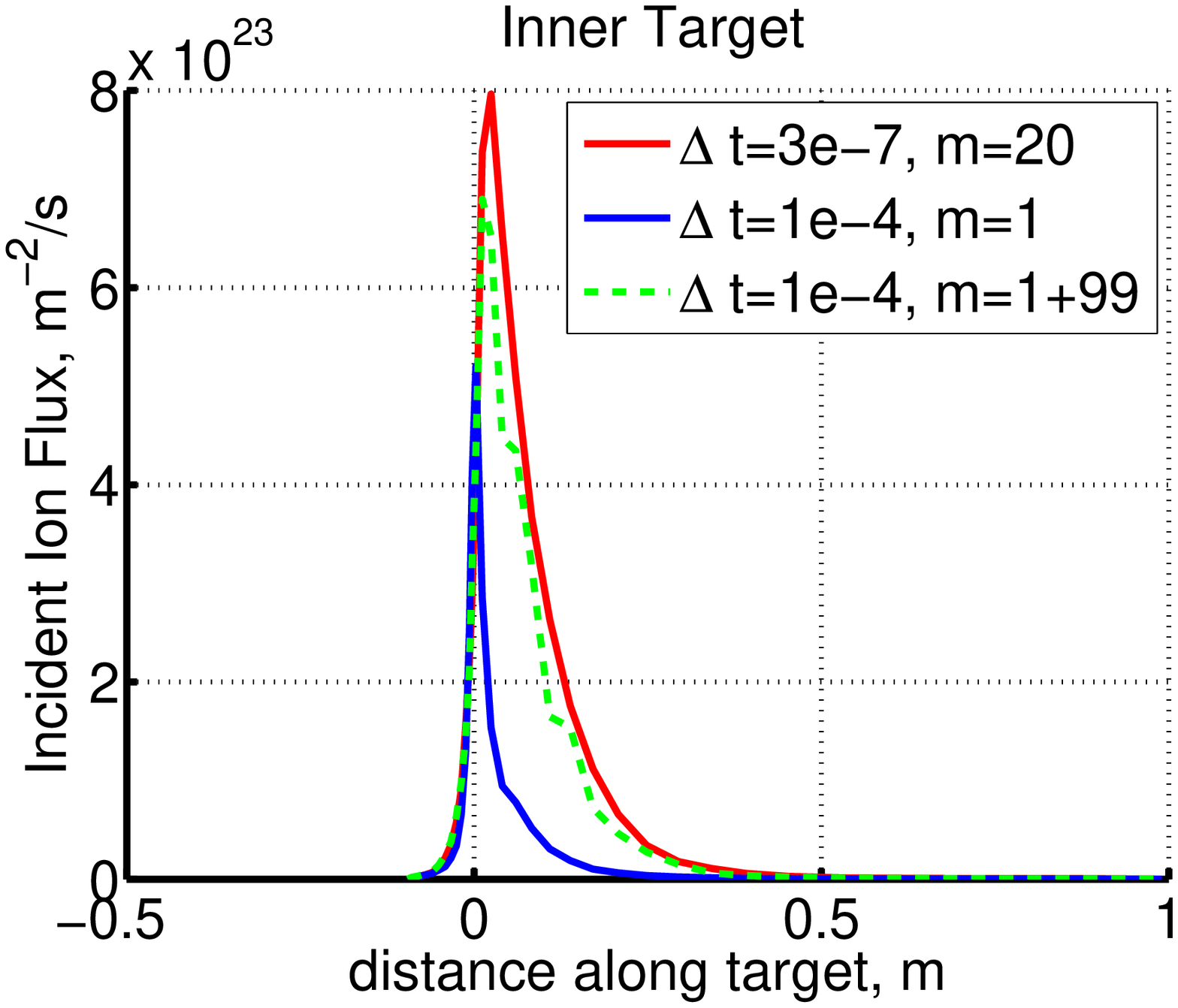}
\includegraphics[width=5cm]{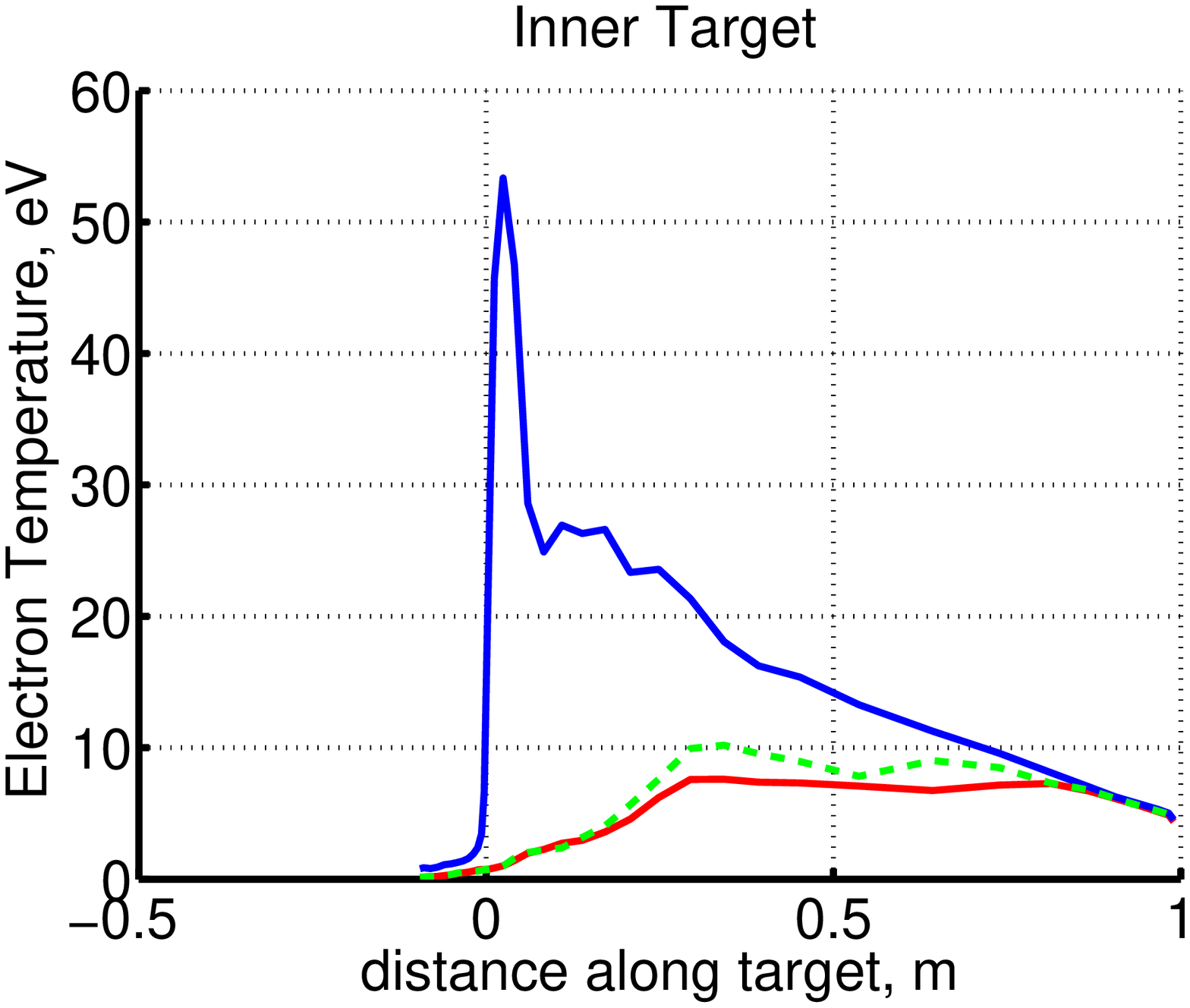}
\includegraphics[width=5cm]{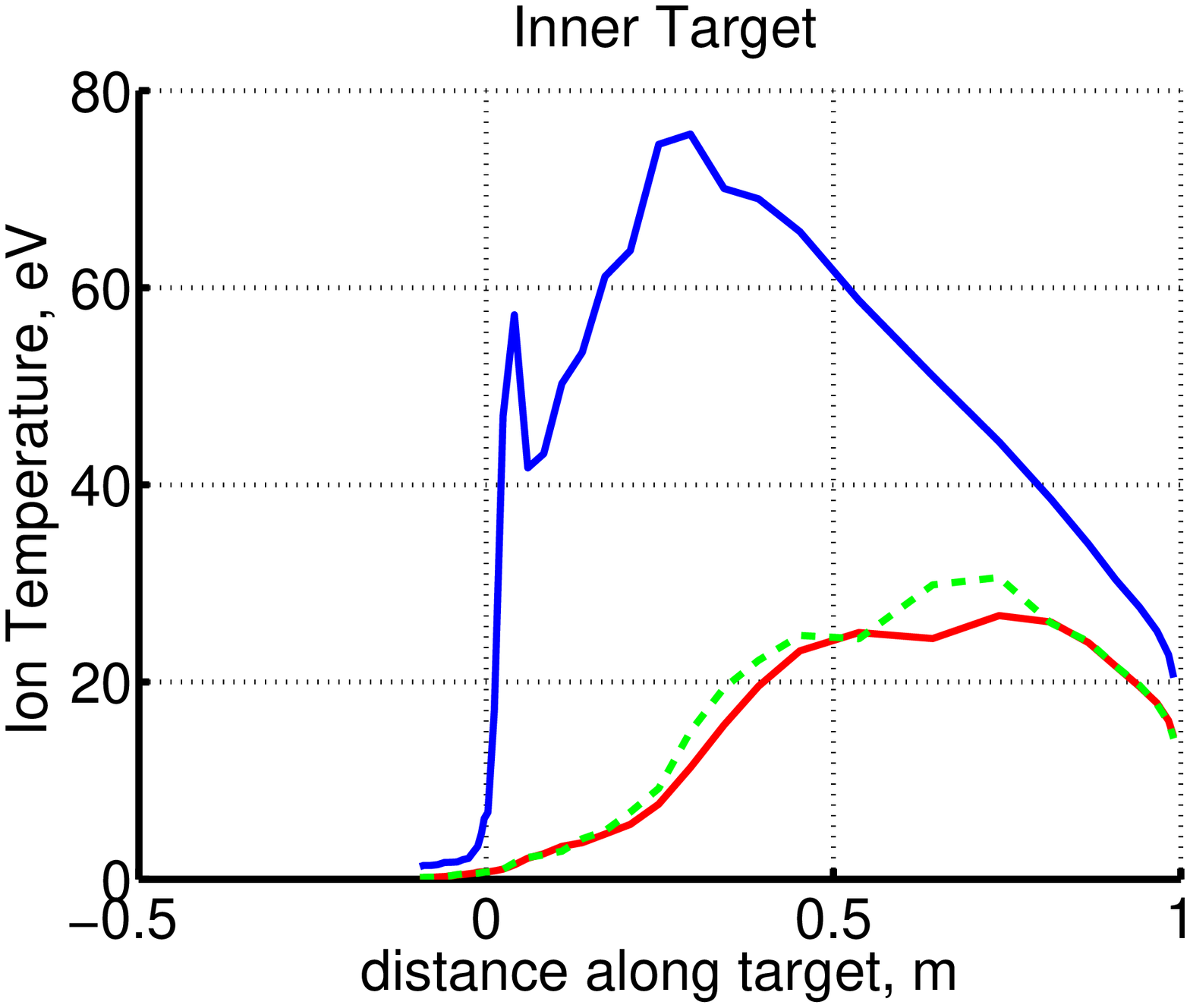} \\ 
\includegraphics[width=5cm]{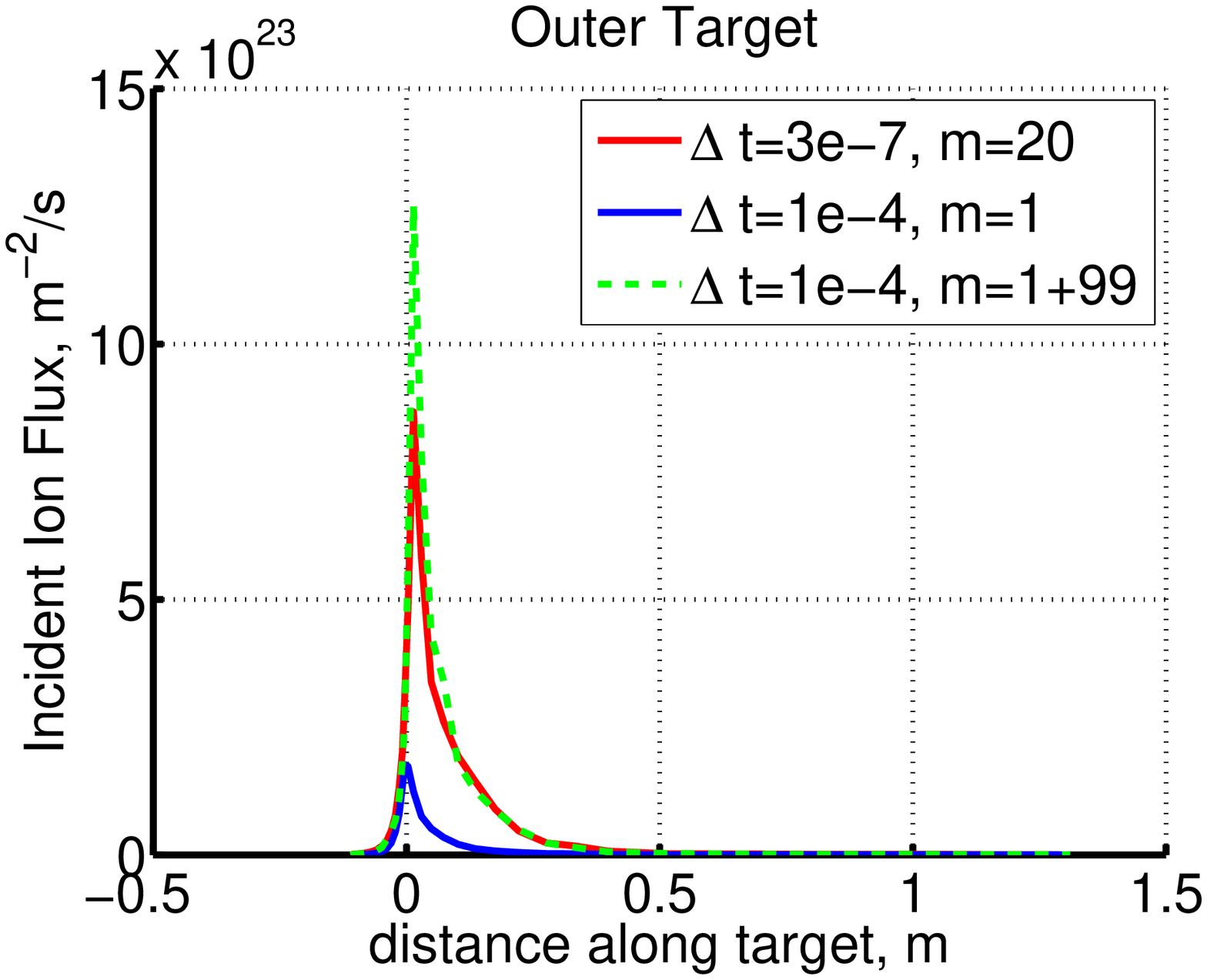}
\includegraphics[width=5cm]{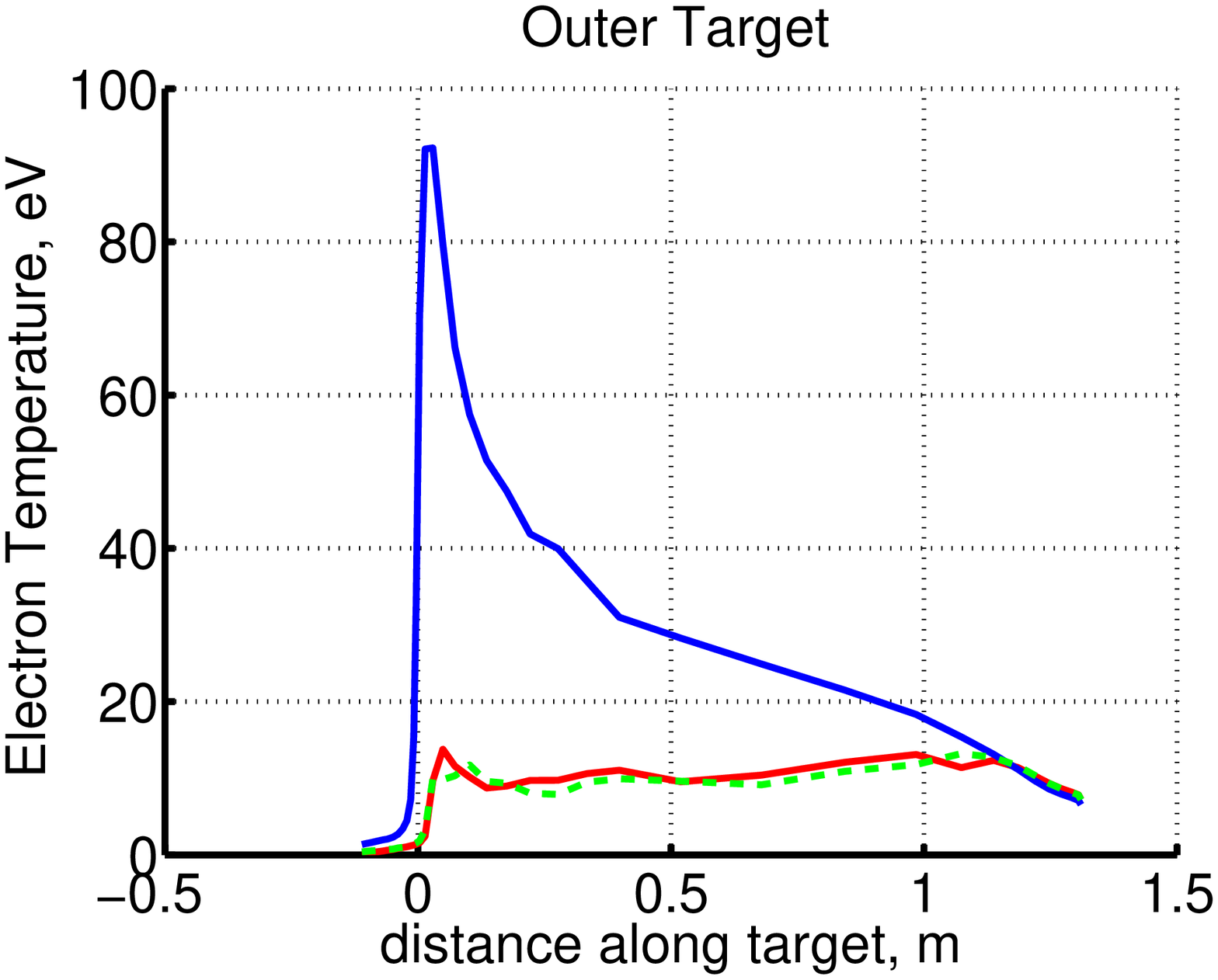}
\includegraphics[width=5cm]{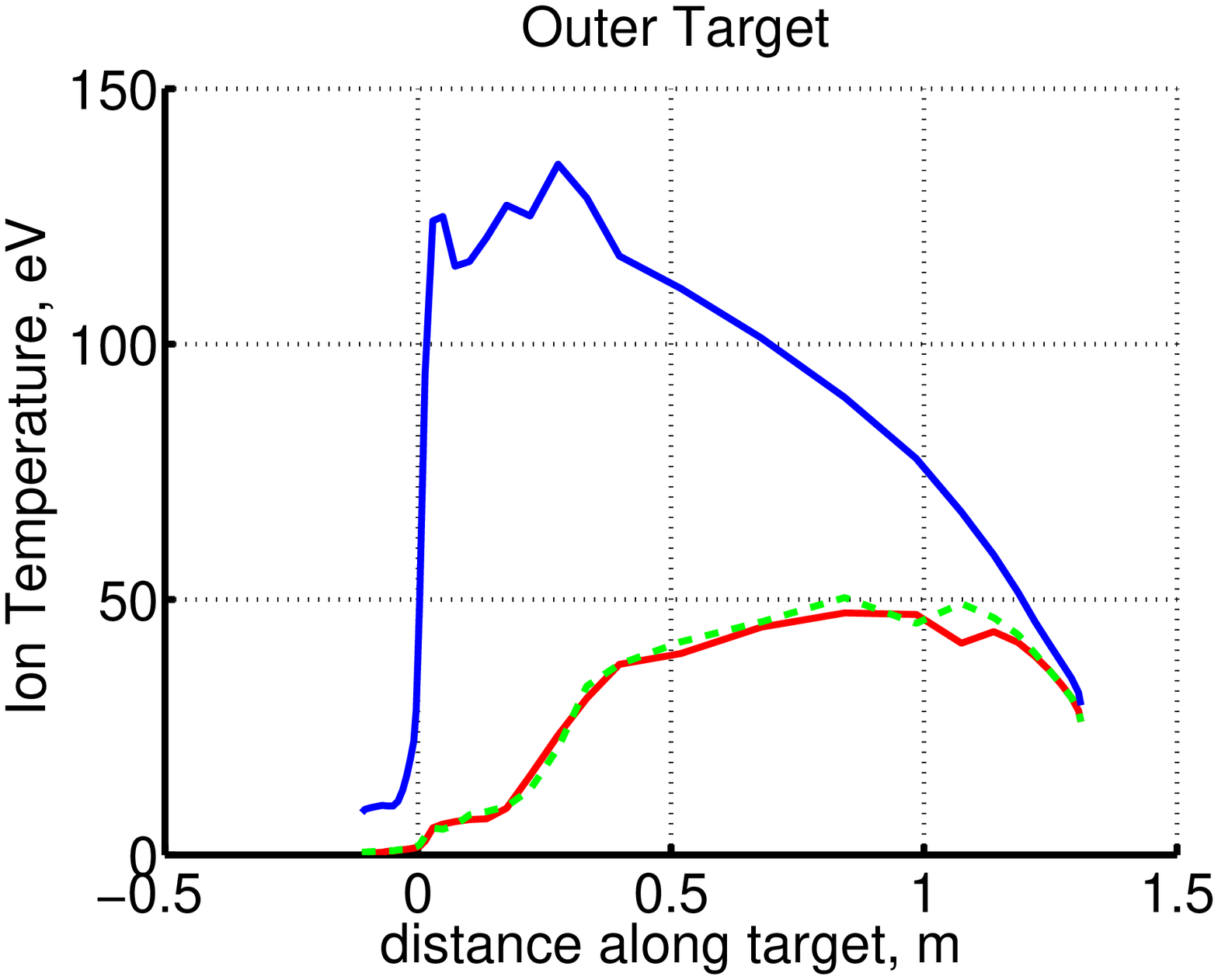} \\
\caption{B2-EIRENE solutions for ITER with small ($\Delta t$=3e-7, $m$=20) and large ($\Delta t$=1e-4, $m$=1) error in the global particle balances, 
         Section~\ref{exampleS}, obtained subsequently with and without internal iterations after each call of the Monte-Carlo model for 
         neutral particles. 
         Parameters in the first ring outside separatrix are plotted from X-point to X-point.
         Inner divertor throat is on the left. 
         In the target plots zero distance is the separatrix, negative coordinates are in the Private Flux Region.
         Dashed line ($\Delta t=$1e-4, $m$=1+99) is the solution obtained with extra iterations 
         for continuity equations only, see Section~\ref{reductionS}}
\label{comparing_solutionsF}        
\end{figure*}

In the ITER modelling studies~\cite{KukushaFED2011,KukushaPPCF2002,KukushaNF2009} 
the B2-EIRENE code was always applied with internal iterations in the 
fluid solver. In the model run in question $m$=20 internal iterations are 
used, the time-step is $\Delta t$=3e-7~sec. Significant increase of the time-step  
is not possible: with $\Delta t>$1e-6~sec a numerical instability 
develops and no stationary solution can be found. 
It turns out that $\Delta t$ 
can be increased by orders of magnitude if no internal iterations are applied, that is
$m=1$. In this case no visible instability develops even with $\Delta t$=1e-4~sec.
However, solutions obtained with and without internal iterations - they are shown in Figure~\ref{comparing_solutionsF} -
strongly deviate from each other.

Strictly-speaking, in the presence of Monte-Carlo noise in the source terms the solutions 
never reach true steady-state. One can only speak about quasi-steady-state solution which 
randomly oscillates around some average. As applied to the B2-EIRENE runs, the ``quasi-steady-state'' 
is defined through characteristic decay times of selected parameters derived from their time-traces, 
see Appendix. In practice the run is regarded as converged if the condition of quasi-steady-state is fulfilled, 
and if errors in the global power and particle balances are small.

Errors in the balances for the two model runs considered here are given in Table~\ref{particle_energy_balanceT}.
The power balance error $\Delta P$ is defined by Equation~(\ref{energy_balance_diagEq}),   
$\Delta \Gamma$ are calculated using Equation~(\ref{balance_fluxesEq}).
One can see that $\Delta P$ is small in both cases. The situation is completely different 
for the particle balance. Whereas in the simulation made with $m=20$ both $\Delta \Gamma^{D,He}$$<$10~\%, 
in the ``fast'' run the error approaches 100~\%. 
That is, in the solution obtained with $m=1$ the pumped fluxes are negligible compared to $\Gamma^{D,He}_{in}$. 

Individual terms of the error are shown in Table~\ref{particle_balanceT} for both recycling species D and He. 
There is a good agreement between  $\Delta \Gamma$ and $\Delta_R+\Delta_S+\Delta_T$ calculated independently by two diagnostics. 
From Table~\ref{particle_balanceT} the reason of the large error  in $m=1$ case immediately becomes clear.
While $\Delta^D_S$ and $\Delta^D_T$ always remain relatively small, $\Delta_R$ becomes very large if 
the code is operated without internal iterations after each Monte-Carlo call.

Particle balance is much more 
difficult to converge than the power balance because of different relation between the 
controlling flux and internal sources and sinks in the system. For the power the 
sources and sinks in plasma are smaller than $P_{SOL}$. In contrast, the particles 
are ``recycled'' between the plasma and solid surfaces and the total volumetric ion 
sources by far exceed $\Gamma_{in}$. In the present example 
$\sum_i \tilde S^{D}$=4.3e24~s$^{-1}$ and $\sum_i \tilde S^{He}$=8.1e22~s$^{-1}$. 
Those numbers are by more than two orders of magnitude larger than  $\Gamma_{in}$ of those 
species. This problem does not appear for C because in the present model all incident C
particles are absorbed on the surfaces - this species does not recycle.

\begin{table}
\caption{Relative errors of the particle and power balances in the model runs of Section~\ref{exampleS}
}
\begin{tabular}{l|cccc}
\hline
    case                        & $\Delta P, \%$  & $\Delta \Gamma^D, \%$ & $\Delta \Gamma^{He}, \%$ & $\Delta \Gamma^{C}, \%$ \\
\hline
 $m$=20,  $\Delta t$=3e-7       &     0.74        &        1.39            &          6.77           &   0.023  \\
 $m$=1, $\Delta t$=1e-4         &     0.32        &        91.5            &          99.3           &   4.6    \\
 $m$=1+99, $\Delta t$=1e-4$^*$  &     1.8         &        16.1            &          14.0           &   0.53   \\
\hline
\multicolumn{5}{l}{$^*$this case is discussed in Section~\ref{reductionS}}
\end{tabular}
\label{particle_energy_balanceT}
\end{table}

\begin{table}
\caption{Individual terms of the error in particle balance (in \%), Equation~(\ref{balance_residualEq}), 
         model runs of Section~\ref{exampleS}}
\begin{tabular}{l|ccc|ccc}
\hline  
    case                      & $\Delta^D_R$ &  $\Delta^D_S$ & $\Delta^D_T$  & $\Delta^{He}_R$ &  $\Delta^{He}_S$ & $\Delta^{He}_T$  \\
\hline  
$m$=20, $\Delta t$=3e-7       &   0.02       &    0.14       &    -1.54      &     3.10        &     0.67         &    -10.63         \\
$m$=1,  $\Delta t$=1e-4       &   91.6       &    1.06       &    -1.98      &     100.7       &     -0.11        &     0.15         \\
$m$=1+99, $\Delta t$=1e-4$^*$ &   0.17       &    16.3       &     0.19       &    -0.10       &      4.52        &     8.92         \\
\hline  
\multicolumn{7}{l}{$^*$this case is discussed in Section~\ref{reductionS}}
\end{tabular}
\label{particle_balanceT}
\end{table}

In the B2-EIRENE model run above extra measures for reduction of residuals on each time-iteration were absolutely 
necessary. Only in this case a solution can be obtained which is correct in terms of the global balances.
The techniques such as internal iterations in B2 can impose severe limitation on the time-step, and it is not attractive 
from the run-time point of view to operate the code in this mode. 
Experience has shown that the use of B2-EIRENE with $m$=1 does not always lead to deviations as dramatic as that shown in 
Figure~\ref{comparing_solutionsF}. E.g. in ITER cases with single fluid (D only) plasma $\Delta \Gamma^D$ was found to 
be sufficiently small both with and without internal iterations, and the obtained solutions are close to each other, 
see example in Ref.~\onlinecite{KotovJul2014}, Chapter 4.1.

The multi-fluid simulation analyzed here   
clearly demonstrates that this must not always be the case. This example emphasizes that in each simulation the particle balance 
has to be carefully monitored with the special diagnostics. Too large error detected by the diagnostic is an unequivocal indication that the residual 
reduction techniques must be applied irrespective of the run-time penalty which they impose.

\section{Reduction of residuals}

\label{reductionS}

A series of studies was undertaken with the B2-EIRENE code 
to find algorithms which would deliver sufficiently good accuracy without 
penalizing the run-time. Their outcome may be of general interest  
for developers and users of other edge modelling codes as well. 
Main results are briefly summarized in this section. 

As a simplest remedy to the particle balance problem a ``0D correction'' 
was first tried, see Ref.~\onlinecite{KotovJul2014}, Chapter 5.5. The 
ion density in the whole computational domain is multiplied by
a constant factor calculated in such way that with the corrected ion fluxes
$\Delta_R$ automatically becomes zero. It was found that this method 
cannot be used because it always produces solutions oscillating in time, and no stationary solutions.

Much more success was achieved with a correction based on iterative relaxation 
of the finite-volume continuity equations.
Technical details of the implementation in B2 can be found in Ref.~\onlinecite{KotovJul2014}, Chapter 5.2-5.4. 
This algorithm works as follows. The whole set of equations for particle, momentum and 
energy balances is relaxed only on the first internal iteration. On subsequent 
iterations only equations for particle continuity are relaxed. To be precise, in 
the code B2 those are pressure correction equations where both the density and velocity fields 
are modified. (B2 uses compressible version of the Patankar's SIMPLE algorithm,
see Ref.~\onlinecite{PatankarBook1980}, Chapter 6.7 and Ref.~\onlinecite{BraamsNET}, Chapter 3.)
Nevertheless, correction of the particle balance via relaxation of the pressure correction equations 
was found to be very reliable. Tests performed for the same ITER model as in 
Section~\ref{exampleS} showed that such iterations robustly converge with time-steps up to $\Delta t$=1e-4~sec.

Results obtained with this algorithm  can be found in the last row in
Tables~\ref{particle_energy_balanceT} and~\ref{particle_balanceT}. The run was performed with 
99 iterations for continuity equations after one full internal iteration, which is
reflected in the designation $m$=1+99. Despite increased $\Delta_S$ the method leads to significant reduction 
of the total error $\Delta$ due to reduction of $\Delta_R$. 
As expected, the main disadvantage of this procedure is that it increases 
residuals of other equations. Closer investigations (Ref.~\onlinecite{KotovJul2014}, Chapter 6.1) showed that 
especially the parallel momentum balance suffers. However, comparison of the solutions obtained with full internal 
iterations and with the reduced scheme demonstrates that they are close to each other: 
the $m$=1+99 case is shown by dashed lines in Figure~\ref{comparing_solutionsF}. Moreover, 
the tests demonstrated that this result holds even for the ITER model with 
high density detached divertor, see Ref.~\onlinecite{KotovJul2014}, Chapter 6.4.
Hence, the method can be suggested for use in a two stage approach for fast
finding of the initial approximation to the solution which is then refined on the 
second ``slow'' stage by more accurate techniques. 

As a next step a scheme was proposed  where coupled continuity and 
parallel momentum balance equations are iterated  - without equations for temperatures. 
This kind of ``incomplete internal iterations'' was implemented in B2-EIRENE and tested as well, 
but the results were found to be unsatisfactory: Ref.~\onlinecite{KotovApr2015}, Chapter 2.3. Tests showed that similar to the 
full internal iterations the ``incomplete iterations'' are prone to numerical instabilities with 
large time-steps, and therefore bring no advantages. The SIMPLE pressure correction which introduces 
extra non-linearity is a possible reason of this behavior. The scheme could be improved if 
monolithic coupling of the continuity and momentum equation would be applied instead. 
That is, when corrections for both the density and the velocity fields are calculated simultaneously 
in a one set of linear equations. 

A fairly simple technique which increases accuracy and can be easily implemented in any code 
is time-averaging of source terms, Ref.~\onlinecite{KotovJul2014}, Chapter 3. 
Although this algorithm can be helpful in many cases, 
it was found to be not always efficient enought in reducing $\Delta \Gamma$,  
in particular with impurities, see example in Ref.~\onlinecite{KotovJul2014}, Chapter 4.2.
In~\cite{KawashimaPFR2006} a more advanced ``piling method" is described which do not 
reset the whole history as the calculation of the new average starts. 

Finally, the brute force method can always be applied to decrease both the statistical error in the 
source terms and the residuals - massive increase of the number of test particles. 
Applicability of this solution strongly depends on the available computing 
hardware. The test particle Monte-Carlo algorithm is easy to parallelize, and  
the increased number of particles does not necessarily mean the increased wall-clock run time.
Experience~\cite{KotovJul2014} has indicated that the pure ``brute force 
compensation'' of the particle balance issue described in Section~\ref{exampleS} is likely 
to require $>>$100 processors to be practical. 

\section{Conclusions}

Use of the test particle Monte-Carlo for neutrals in the tokamak edge modelling codes 
has an unpleasant side effect of random error in the source terms. 
If no special measures are taken, then this persistent statistical noise leads 
to residuals of the discretized fluid equations which do not converge, but saturate at a certain level. 
In the present paper one particular well identified issue caused by the saturated residuals 
has been described. It has been shown that too large finite-volume residuals can cause crude violation 
of the global particle balance. In turn, for the system in question - the tokamak edge and divertor plasma - 
violation of the particle conservation may have a very strong (``zero order'') non-local impact on the whole numerical solution.

There are computational techniques which can effectively reduce the residuals. 
E.g. in the code B2 which uses splitting by equations an extra loop of simple iterations  on 
each time-iteration is applied. However, severe restriction imposed by those internal iterations on the time-step 
leads to a very long overall model run-time when this option is used. 
With numerical diagnostics proposed in this work it can be unambiguously identified when the too large error 
in the particle balance is caused by the saturated residuals, and the residual reduction 
techniques must be applied to obtain the physically meaningful solution. 
The diagnostics can be implemented in any finite-volume edge code. 

The problem describe here would become less of an issue 
if solving the set of non-linear equations on each time-iteration 
would not require reduced time-step. If such solvers are not 
feasible, then the accuracy and run-time drawbacks may even 
outweight the very advantage of using the kinetic test particle Monte-Carlo in the 
self-consistent models. 
The drawbacks can be partly compensated by reducing the statistical error
which is, in principle, only a matter of available computing resources. 
Emerging heterogeneous CPU-booster architectures~\cite{DEEP} could be particularly well suited for 
the combination of a fluid and a Monte-Carlo code. While the serial finite-volume 
part runs on CPU, the Monte-Carlo part can make use of massive parallelization 
on hundreds of processing units on the accelerator. 

\begin{acknowledgments}
This work was performed under EFDA Work Programme 2013 ``Assessment Studies for SOLPS Optimisation'' (WP13-SOL).
\end{acknowledgments}

\appendix

\section{Practical convergence criteria applied to the tokamak edge modelling code B2-EIRENE}

\label{appendixS}

Characteristic time-scale $\tau_X$ of the parameter $X$ is calculated from its 
time-trace $X(t_k)$ by fitting it with a linear function:  
$$
\ln X= \tau^{-1}_X t + C \quad \Rightarrow \quad \frac{1}{X}\frac{d X}{dt} = \frac{1}{\tau_X}
$$
In the present paper the number of last time-iterations used for the fit 
was equal to $\max{\left( 2000, N_p^{(5\;\mu s)}\right)}$, 
where $N_p^{(5\;\mu s)}$ is the number of points which cover last 5~$\mu s$ of physical time.
Least-square method is applied to find the parameters  $\tau_X$ and $C$.
Same data-points were used to calculate average $\Delta \Gamma$ and $\Delta P$
in Table~\ref{particle_energy_balanceT} and $\Delta_{R,S,T}$ in Table~\ref{particle_balanceT}. 

The control parameters for which $\tau_X$ are calculated are the total amount of ions 
$N_\beta$ of species $\beta$, total diamagnetic energy in electrons $E_e$ and ions $E_i$: 
$$
N_\beta=\int \sum_{\alpha'} n_{\alpha'} dV,\quad E_e=\frac32\int n_e T_edV
$$
$$
E_i=\int \left(\frac32\sum_\alpha n_\alpha T_i + \frac12\sum_\alpha m_\alpha n_\alpha v^2_\alpha \right) dV
$$
as well as plasma parameters averaged along the magnetic separatrix: 
$<n_e>^{sep}$, $<T_e>^{sep}$, $<T_i>^{sep}$.
Here the integration is performed over the whole computational grid, $V$ is geometrical volume, 
$\sum_{\alpha}$ is the sum over all ion fluids,
$n_e$ is the electron density, $m_\alpha$ is the atomic mass of ions, $v_\alpha$ is their 
average macroscopic velocity.
The B2-EIRENE solutions analyzed in this paper were regarded as stationary when $\tau_X>$3~sec 
for all the parameters listed above. For $N_D$ and $N_{He}$ $\tau_X>$15~sec. 

Besides this condition of steady-state the errors in the global particle and 
power balances are checked. The error in particle balance is 
expressed by Equation~(\ref{balance_fluxesEq}). Relative error in the power balance 
is defined as follows:
\begin{equation}
\Delta P = \frac{ P_{SOL} - P^+_{PFC} -  P^n_{PFC} - P_{rad} - P^n_{core} }{P_{SOL}}
\label{energy_balance_diagEq}
\end{equation}
Here $P_{SOL}$ is the power influx into the computational domain from the core plasma, 
$P^+_{PFC}$ is the power deposited by charged particles to the Plasma Facing Components (PFC),
$P^n_{PFC}$ is the power deposited to PFC by neutrals, 
$P_{rad}$ is the power radiated by both charged and neutral particles,
$P^n_{core}$ is the power transferred by neutrals back to the core. 


\end{document}